\documentclass[preprintnumbers,amsmath,amssymbm,prl]{revtex4}
\usepackage{epsfig}
\usepackage{graphicx}

\begin{document}
\title{Slow relaxation of rapidly rotating black holes}
\author{Shahar Hod}
\address{The Ruppin Academic Center, Emeq Hefer 40250, Israel}
\address{ }
\address{The Hadassah Institute, Jerusalem 91010, Israel}
\date{\today}

\begin{abstract}
\ \ \ We study {\it analytically} the relaxation phase of perturbed,
rapidly rotating black holes. In particular, we derive a simple
formula for the fundamental quasinormal resonances of near-extremal
Kerr black holes. The formula is expressed in terms of the
black-hole physical parameters: $\omega=m\Omega-i2\pi T_{BH}(n+{1
\over 2})$, where $T_{BH}$ and $\Omega$ are the temperature and
angular velocity of the black hole, and $m$ is the azimuthal
harmonic index of a co-rotating equatorial mode. This formula
implies that the relaxation period $\tau\sim 1/\Im\omega$ of the
black hole becomes extremely long as the extremal limit $T_{BH}\to
0$ is approached. The analytically derived formula is shown to agree
with direct numerical computations of the black-hole resonances. We
use our results to demonstrate analytically the fact that
near-extremal Kerr black holes saturate the recently proposed
universal relaxation bound.
\end{abstract}
\bigskip
\maketitle


The radiative perturbations of a complete gravitational collapse
decay with time leaving behind a `bald' black hole. This is the
essence of the no-hair conjecture introduced by Wheeler
\cite{Wheeler} more than thirty years ago. It asserts that
perturbation fields left outside the collapsing star would either be
radiated away to infinity, or be swallowed by the newly born black
hole.

According to the uniqueness theorems \cite{un1,un2,un3,un4,un5}, the
metric outside the black hole should relax into a Kerr-Newman
spacetime, characterized solely by the black-hole mass, charge, and
angular momentum. This relaxation phase in the dynamics of perturbed
black holes is characterized by `quasinormal ringing', damped
oscillations with a discrete spectrum (see e.g. \cite{Nollert1} for
a detailed review). At late times, all perturbations are radiated
away in a manner reminiscent of the last pure dying tones of a
ringing bell \cite{Press,Cruz,Vish,Davis}.

The black hole quasinormal modes (QNMs) correspond to solutions of
the perturbations equations with the physical boundary conditions of
purely outgoing waves at spatial infinity and purely ingoing waves
crossing the event horizon \cite{Detwe}. Such boundary conditions
single out a {\it discrete} set of black-hole resonances
$\{\omega_n\}$ (assuming a time dependence of the form $e^{-i\omega
t}$). In analogy with standard scattering theory, the QNMs can be
regarded as the scattering resonances of the black-hole spacetime.
They thus correspond to poles of the transmission and reflection
amplitudes of a standard scattering problem in a black-hole
spacetime.

The characteristic quasinormal frequencies are {\it complex}. This
reflects the fact that the perturbations fields decay with time, in
accord with the spirit of the no-hair conjecture. It turns out that
there exist an infinite number of quasinormal modes, characterizing
oscillations with decreasing relaxation times (increasing imaginary
part) \cite{Leaver}. The mode with the smallest imaginary part
(known as the {\it fundamental} mode) determines the characteristic
dynamical timescale $\tau$ for generic perturbations to decay.

Quasinormal resonances are expected to play a prominent role in
gravitational radiation emitted by a variety of astrophysical
scenarios involving black holes. Being the characteristic `sound' of
the black hole itself, these free oscillations are of great
importance from the astrophysical point of view. They allow a direct
way of identifying the spacetime parameters, especially the mass and
angular momentum of the black hole. This has motivated a flurry of
research during the last four decades aiming to compute the
quasinormal mode spectrum of various types of black-hole spacetimes
\cite{Nollert1}.

It is well known that realistic stellar objects generally rotate
about their axis, and are therefore not spherical. Thus, an
astrophysically realistic model of wave dynamics in black-hole
spacetimes must involve a non-spherical background geometry with
angular momentum. In this work we determine analytically the
fundamental (least-damped) resonant frequencies of such
rapidly-rotating Kerr black holes. (For a recent progress in the
study of the highly-damped resonances, see \cite{KeshHod,Ber3}.) The
spectrum of quasinormal resonances can be studied analytically in
the near-extremal limit $(M^2-a^2)^{1/2}\ll a\lesssim M$, where $M$
and $a$ are the mass and angular momentum per unit mass of the black
hole, respectively.

Before going on we would like to summarize what is already known
about these fundamental, slowly-damped Kerr QNMs:

\begin{itemize}
\item In the extremal limit ($a\to M$) one finds $\Re\omega \simeq m/2M$, where $m>0$ is the azimuthal harmonic index of
a perturbation field co-rotating with the black hole. This fact has
been found in many numerical computations (see e.g.,
\cite{Leaver,Ono,GlaAnd2int,GlaAnd1}) and is well understood
analytically \cite{Det,GlaAnd2,Car}.

\item Numerical computations \cite{Leaver,Ono,GlaAnd2int,GlaAnd1} have indicated
that, for co-rotating modes $\Im\omega \to 0$ in the extremal limit.
This implies that these modes are long lived. Detweiler \cite{Det}
presented a semi-analytical formula for the long-lived black-hole
resonances in the extremal limit. It is important to emphasize that
Detweiler's result assumes that $\Im\omega/T_{BH}>>1$ and is not
appropriate for the most long-lived modes in the non-extremal case,
where it might be that $\Im\omega \sim T_{BH}$, where
$T_{BH}={{(M^2-a^2)^{1/2}}\over{4\pi M[M+(M^2-a^2)^{1/2}]}}$ is the
Bekenstein-Hawking temperature of the black hole.

\item Motivated by the recently proposed time-temperature universal relaxation bound (TTT)
\cite{Hod1}, we have recently re-examined the role played by the
imaginary parts of the black-hole quasinormal resonances. It has
been observed \cite{Hod1,Hod2} that the {\it numerically} computed
equatorial resonances of near-extremal Kerr black holes are
well-approximated by the simple {\it analytical} relation
$\omega=m\Omega-i2\pi T_{BH}(n+1/2)$, where $n=0,1,2,...$. The
successfulness of this conjectured formula is demonstrated in Table
\ref{Table1}. The numerical results therefore imply
$\Im\omega=O(T_{BH})$ in the near-extremal limit. A similar relation
which is valid only for the fundamental mode was obtained in
\cite{Gruz} for equatorial modes. Table \ref{Table2} demonstrates
the fact that the predictions of the formula improve as the extremal
limit is approached.
\end{itemize}

\begin{table}[htbp]
\centering
\begin{tabular}{|c|c|c|c|}
\hline
$n$ & $\Re\omega$ & $\Im\omega$ (numerical) & $\Im\omega$ (analytical)\   \\
\hline
\ 0\ \ & \ 0.99324\ \ &\ 0.00341\ \ & 0.00348 \\
\ 1\ \ & \ 0.99322\ \ &\ 0.01020\ \ & 0.01045 \\
\ 2\ \ & \ 0.99321\ \ &\ 0.01699\ \ & 0.01743 \\
\ 3\ \ & \ 0.99320\ \ &\ 0.02385\ \ & 0.02440 \\
\ 4\ \ & \ 0.99317\ \ &\ 0.03067\ \ & 0.03137 \\
\ 5\ \ & \ 0.99313\ \ &\ 0.03749\ \ & 0.03834 \\
\hline
\end{tabular}
\caption{Quasinormal resonances of a near-extremal Kerr black hole
with $a/M=0.9999$. The data shown is for the equatorial mode
$l=m=2$, see also \cite{GlaAnd2int}. The proposed analytical formula
is $\omega=m\Omega-i2\pi T_{BH}(n+1/2)$. (Here $m\Omega\simeq
0.9859$). The agreement between the numerical data and the proposed
formula is of $\sim 2\%$.} \label{Table1}
\end{table}

\begin{table}[htbp]
\centering
\begin{tabular}{|c|c|c|c|}
\hline
$a/M$ & $\Re\omega_{ana}/\Re\omega_{num}$ & $\Im\omega_{ana}/\Im\omega_{num}$ \\
\hline
\ 0.9\ \ & \ 0.933\ \ &\ 1.170\ \\
\ 0.96\ \ & \ 0.977\ \ &\ 1.106\ \\
\ 0.9999\ \ & \ 0.993\ \ &\ 1.022\ \\
\hline
\end{tabular}
\caption{The ratio between the relation $\omega=m\Omega-i\pi T_{BH}$
for the fundamental equatorial black-hole resonance, and the
numerically computed value. The data shown is for the mode $l=m=2$.
The agreement between the numerical data and the analytical formula
improves as the black hole approaches its extremal limit.}
\label{Table2}
\end{table}

In order to determine the black-hole resonances we shall analyze the
scattering of massless waves in the Kerr spacetime. The dynamics of
a perturbation field $\Psi$ in the rotating Kerr spacetime is
governed by the Teukolsky equation \cite{Teu}. One may decompose the
field as (we use natural units in which $G=c=\hbar=1$)
\begin{equation}\label{Eq1}
\Psi_{slm}(t,r,\theta,\phi)=e^{im\phi}S_{slm}(\theta;a\omega)\psi_{slm}(r)e^{-i\omega
t}\ ,
\end{equation}
where $(t,r,\theta,\phi)$ are the Boyer-Lindquist coordinates,
$\omega$ is the (conserved) frequency of the mode, $l$ is the
spheroidal harmonic index, and $m$ is the azimuthal harmonic index
with $-l\leq m\leq l$. The parameter $s$ is called the spin weight
of the field, and is given by $s=\pm 2$ for gravitational
perturbations, $s=\pm 1$ for electromagnetic perturbations, $s=\pm
{1\over 2}$ for massless neutrino perturbations, and $s=0$ for
scalar perturbations. (We shall henceforth omit the indices $s,l,m$
for brevity.) With the decomposition (\ref{Eq1}), $\psi$ and $S$
obey radial and angular equations, both of confluent Heun type
\cite{Heun,Flam}, coupled by a separation constant $A(a\omega)$.

The angular functions $S(\theta;a\omega)$ are the spin-weighted
spheroidal harmonics which are solutions of the angular equation
\cite{Teu,Flam}
\begin{equation}\label{Eq2}
{1\over {\sin\theta}}{\partial \over
{\partial\theta}}\Big(\sin\theta {{\partial
S}\over{\partial\theta}}\Big)+\Big[a^2\omega^2\cos^2\theta-2a\omega
s\cos\theta-{{(m+s\cos\theta)^2}\over{\sin^2\theta}}+s+A\Big]S=0\ .
\end{equation}
The angular functions are required to be regular at the poles
$\theta=0$ and $\theta=\pi$. These boundary conditions pick out a
discrete set of eigenvalues $A_l$ labeled by an integer $l$. [In the
$a\omega\ll 1$ limit these angular functions become the familiar
spin-weighted spherical harmonics with the corresponding angular
eigenvalues $A=l(l+1)-s(s+1)+O(a^2\omega^2)$.]

The radial Teukolsky equation is given by
\begin{equation}\label{Eq3}
\Delta^{-s}{{d}
\over{dr}}\Big(\Delta^{s+1}{{d\psi}\over{dr}}\Big)+\Big[{{K^2-2is(r-M)K}\over{\Delta}}
-a^2\omega^2+2ma\omega-A+4is\omega
r\Big]\psi=0\ ,
\end{equation}
where $\Delta\equiv r^2-2Mr+a^2$ and $K\equiv (r^2+a^2)\omega-am$.
The zeroes of $\Delta$, $r_{\pm}=M\pm (M^2-a^2)^{1/2}$, are the
black hole (event and inner) horizons.

For the scattering problem one should impose physical boundary
conditions of purely ingoing waves at the black-hole horizon and a
mixture of both ingoing and outgoing waves at infinity (these
correspond to incident and scattered waves, respectively). That is,
\begin{equation}\label{Eq4}
\psi \sim
\begin{cases}
e^{-i\omega y}+{\mathcal{R}}(\omega)e^{i \omega y} & \text{ as }
r\rightarrow\infty\ \ (y\rightarrow \infty)\ ; \\
{\mathcal{T}}(\omega)e^{-i (\omega-m\Omega)y} & \text{ as }
r\rightarrow r_+\ \ (y\rightarrow -\infty)\ ,
\end{cases}
\end{equation}
where the ``tortoise" radial coordinate $y$ is defined by
$dy=[(r^2+a^2)/\Delta]dr$. Here $\Omega\equiv {a\over {2Mr_+}}$ is
the angular velocity of the black-hole horizon. The coefficients
${\cal T}(\omega)$ and ${\cal R}(\omega)$ are the transmission and
reflection amplitudes for a wave incident from infinity. The
discrete quasinormal frequencies are the scattering resonances of
the black-hole spacetime. They thus correspond to poles of the
transmission and reflection amplitudes. (The pole structure reflects
the fact that the QNMs correspond to purely outgoing waves at
spatial infinity.) These resonances determine the ringdown response
of a black hole to outside perturbations.

The transmission and reflection amplitudes satisfy the usual
probability conservation equation $|{\cal T}(\omega)|^2+|{\cal
R}(\omega)|^2=1$. Teukolsky and Press \cite{TeuPre} and also
Starobinsky and Churilov \cite{StaChu} have analyzed the black-hole
scattering problem in the double limit $a\to M$ and $\omega\to
m\Omega$. Detweiler \cite{Det} then used that solution to determine
the long-lived black-hole resonances in the extremal limit. Define
\begin{equation}\label{Eq5}
\sigma\equiv {{r_+-r_-}\over{r_+}}\ \ ;\ \ \tau\equiv
M(\omega-m\Omega)\ \ ;\ \ \hat\omega\equiv \omega r_+\  .
\end{equation}
Then the resonance condition obtained in \cite{Det} for $\sigma<<1$
and $\tau<<1$ is:
\begin{equation}\label{Eq6}
-{{\Gamma(2i\delta)\Gamma(1+2i\delta)\Gamma(1/2+s-2i\hat\omega-i\delta)\Gamma(1/2-s-2i\hat\omega-i\delta)}\over
{\Gamma(-2i\delta)\Gamma(1-2i\delta)\Gamma(1/2+s-2i\hat\omega+i\delta)\Gamma(1/2-s-2i\hat\omega+i\delta)}}=
(-2i\hat\omega\sigma)^{2i\delta}
{{\Gamma(1/2+2i\hat\omega+i\delta-4i\tau/\sigma)}\over{\Gamma(1/2+2i\hat\omega-i\delta-4i\tau/\sigma)}}\
,
\end{equation}
where $\delta^2\equiv 4\hat\omega^2-1/4-A-a^2\omega^2+2ma\omega$. In
the extremal limit Detweiler solved this equation with the
assumption that $\tau/\sigma\to\infty$. For near extremal black
holes, it should be emphasized that the numerical data presented
above is consistent with the limit $\tau/\sigma\to const$ as
$\sigma\to 0$. Thus the quasinormal frequencies obtained in
\cite{Det} do not include the most long lived, fundamental
resonances of near-extremal black holes. We now derive analytically
the fundamental resonances of rapidly rotating, near-extremal Kerr
black holes.

The left-hand-side of Eq. (\ref{Eq6}) has a well defined limit as
$a\to M$ and $\omega\to m\Omega$. We denote that limit by ${\cal
L}$. In the limit $\omega\to m\Omega$, where $\omega$ is almost
purely real, one finds from Eq. (\ref{Eq2}) that the separation
constants $\{A\}$ are also almost purely real. This in turn implies
that the $\delta^2$'s are almost purely real. For some modes,
including most of the equatorial $l=m$ modes (and also other modes
with $m$ close enough to $l$), $\delta^2$ is found to be positive,
which implies that in these cases $\delta$ is almost purely real and
positive \cite{TeuPre,PreTeu}. For the rest of the modes one finds
$\delta^2<0$, which implies that $\delta$ is almost purely imaginary
with positive imaginary part \cite{TeuPre,PreTeu}.


If $\delta$ is almost purely real and positive then one has
$(-i)^{-2i\delta}=e^{-2i\delta\ln(-i)}=e^{-2i\delta\ln
e^{-i\pi/2}}=e^{-2i\delta(-i\pi/2)}=e^{-\pi\delta}\ll 1$. Here we
have used the fact that $\delta>2$ for all gravitational equatorial
modes \cite{TeuPre,PreTeu}. (In fact, this is also true for many
other modes for which $l\sim m\gg 1$.) If $\delta$ is almost purely
imaginary with a positive imaginary part then one has
$\sigma^{-2i\delta} \to 0$ in the near-extremal limit $\sigma\to 0$.
In both cases one therefore finds
$\epsilon\equiv(-2i\hat\omega\sigma)^{-2i\delta}\ll 1$. Thus, a
consistent solution of the resonance condition, Eq. (\ref{Eq6}), may
be obtained if
$1/\Gamma(1/2+2i\hat\omega-i\delta-4i\tau/\sigma)=O(\epsilon)$
\cite{Notedelta}. Suppose
\begin{equation}\label{Eq7}
1/2+2i\hat\omega-i\delta-4i\tau/\sigma=-n+\eta\epsilon+O(\epsilon^2)\
,
\end{equation}
where $n\geq 0$ is a non-negative integer and $\eta$ is an unknown
constant to be determined below. Then one has
\begin{equation}\label{Eq8}
\Gamma(1/2+2i\hat\omega-i\delta-4i\tau/\sigma)\simeq\Gamma(-n+\eta\epsilon)\simeq
(-n)^{-1}\Gamma(-n+1+\eta\epsilon)\simeq\cdots\simeq [(-1)^n
n!]^{-1}\Gamma(\eta\epsilon)\  ,
\end{equation}
where we have used the relation $\Gamma(z+1)=z\Gamma(z)$
\cite{Abram}. Next, using the series expansion
$1/\Gamma(z)=\sum_{k=1}^{\infty} c_k z^k$ with $c_1=1$ [see Eq.
$(6.1.34)$ of \cite{Abram}], one obtains
\begin{equation}\label{Eq9}
1/\Gamma(1/2+2i\hat\omega-i\delta-4i\tau/\sigma)=(-1)^n
n!\eta\epsilon+O(\epsilon^2)\  .
\end{equation}
Substituting this into Eq. (\ref{Eq6}) one finds $\eta={\cal
L}/[(-1)^n n!\Gamma(-n+2i\delta)]$.

Finally, recalling that $4\tau/\sigma=(\omega-m\Omega)/2\pi T_{BH}$
we obtain from Eq. (\ref{Eq7}) the resonance condition
\begin{equation}\label{Eq10}
(\omega-m\Omega)/2\pi T_{BH}=i[-n+\eta\epsilon-1/2+i(\delta-m)]\  ,
\end{equation}
where we have substituted $2i\hat\omega\simeq im$ for $\omega\simeq
m\Omega\simeq m/2M$. The black-hole quasinormal resonances of
equatorial $l=m\geq 0$ modes (and, in general, co-rotating modes for
which $\delta^2>0$) are therefore given by the leading-order formula
\begin{equation}\label{Eq11}
\omega=m\Omega-i2\pi T_{BH}(n+1/2)+O(T_{BH})\  ,
\end{equation}
where $n=0,1,2,...$ . One thus finds that $\Re\omega\to m\Omega$ and
$\Im\omega\to 0$ in the near-extremal. These analytical findings are
in accord with direct numerical computations, see Tables I and II
\cite{Notevalid}.

The black-hole quasinormal resonances of non-equatorial $l\neq m\geq
0$ modes (and, in general, co-rotating modes for which $\delta^2<0$)
are given by the leading-order formula [see Eq. (\ref{Eq10})]
\begin{equation}\label{Eq12}
\omega=m\Omega-i2\pi T_{BH}(n+1/2-i\delta)+O(T_{BH})\  ,
\end{equation}
where $n=0,1,2,...$. It is worth pointing out that, in this case
$\Im\omega>2\pi T_{BH}(n+1/2)$ (recall that $\Im\delta>0$). This
implies that for these modes $\Im\omega$ approaches zero slower as
compared to the equatorial ones for which $\Im\omega=2\pi
T_{BH}(n+1/2)$. We have therefore established the fact that
non-equatorial modes decay faster than the equatorial ones.

In summary, we have studied analytically the quasinormal spectrum of
rapidly-rotating Kerr black holes. The main results and their
physical implications are as follows:

(1) We have shown that the fundamental resonances can be expressed
in terms of the black-hole physical parameters: the temperature
$T_{BH}$, and the angular velocity $\Omega$ of the horizon.

(2) It was found that for all co-rotating modes (modes having $m>0$)
$\Re\omega\to m\Omega$ in the near-extremal limit. This conclusion
is in agrement with, and generalizes, the $l=m$ result obtained in
\cite{Det}.

(3) We find that, in the near-extremal limit $\Im\omega$ approaches
zero linearly with the black-hole temperature. Namely,
$\Im\omega=O(T_{BH})$. This conclusion holds true for all modes
co-rotating with the black hole. Thus, all co-rotating modes become
long-lived as the black hole spins up. Moreover, it is realized that
equatorial $l=m$ modes (and in general, modes for which the quantity
$\delta$ is real) decay slower than other non-equatorial
perturbations.

(4) It is worth mentioning that a fundamental bound on the
relaxation time $\tau$ of a perturbed thermodynamical system has
recently been suggested \cite{Hod1}, $\tau \geq \hbar/\pi T$, where
$T$ is the system's temperature. Taking cognizance of this
relaxation bound, one deduces an upper bound on the black-hole
fundamental (slowest damped) frequency
\begin{equation}\label{Eq13}
\min\{\Im\omega\} \leq \pi T_{BH}\  .
\end{equation}
Thus the relaxation bound implies that a black hole must have (at
least) one quasinormal resonance whose imaginary part conform to the
upper bound (\ref{Eq13}). This mode would dominate the relaxation
dynamics of the perturbed black hole and will determine its
characteristic relaxation timescale. Taking cognizance of Eq.
(\ref{Eq11}) for the equatorial modes, and substituting $n=0$ for
the fundamental resonance, one obtains $\min\{\Im\omega\}=\pi
T_{BH}+O(T_{BH}^2)$ \cite{Hod1,Hod2,Gruz,Hodb,Notedim}. One
therefore concludes that rapidly rotating Kerr black holes actually
{\it saturate} the universal relaxation bound.

\bigskip
\noindent
{\bf ACKNOWLEDGMENTS}
\bigskip

This research is supported by the Meltzer Science Foundation. I
thank Uri Keshet, Yael Oren, Liran Shimshi and Clovis Hopman for
helpful discussions. I also thank Andrei Gruzinov for interesting
correspondence. It is my pleasure to thank Steve Detweiler for
valuable comments.



\begin{thebibliography}{99}

\bibitem{Wheeler} R. Ruffini and J. A. Wheeler, Physics Today {\bf 24}, 30 (1971).

\bibitem{un1} W. Israel, Phys. Rev. {\bf 164}, 1776 (1967); Commun.
Math. Phys. {\bf 8}, 245 (1968).

\bibitem{un2} B. Carter, Phys. Rev. Lett. {\bf 26}, 331 (1971).

\bibitem{un3} S. W. Hawking, Commun. Math. Phys. {\bf 25}, 152 (1972).

\bibitem{un4} D. C. Robinson, Phys. Rev. D {\bf 10}, 458 (1974); Phys. Rev.
Lett. {\bf 34}, 905 (1975).

\bibitem{un5} J. Isper, Phys. Rev. Lett. {\bf 27}, 529 (1971).

\bibitem{Nollert1} For an excellent review and a detailed list of references see
H. P. Nollert, Class. Quantum Grav. {\bf 16}, R159 (1999).

\bibitem{Press} W. H. Press, Astrophys. J. {\bf 170}, L105 (1971).

\bibitem{Cruz} V. de la Cruz, J. E. Chase and W. Israel,
  Phys. Rev. Lett. {\bf 24}, 423 (1970).

\bibitem{Vish} C.V. Vishveshwara, Nature {\bf 227}, 936 (1970).

\bibitem{Davis} M. Davis, R. Ruffini, W. H. Press and R. H. Price,
  Phys. Rev. Lett. {\bf 27}, 1466 (1971).

\bibitem{Detwe} S. L. Detweiler, in Sources of Gravitational
  Radiation, edited by L. Smarr (Cambridge University Press,
  Cambridge, England, 1979).

\bibitem{Leaver} E. W. Leaver, Proc. R. Soc. A {\bf 402}, 285 (1985).

\bibitem{KeshHod} U. Keshet and S. Hod, Phys. Rev. D {\bf 76}, R061501
(2007).

\bibitem{Ber3} E. Berti, V. Cardoso and S. Yoshida, Phys. Rev. D {\bf
69}, 124018 (2004).

\bibitem{Ono} H. Onozawa, Phys. Rev. D {\bf 55}, 3593 (1997).

\bibitem{GlaAnd2int} See Table I of K. Glampedakis and N. Andersson,
arXiv:gr-qc/0103054.

\bibitem{GlaAnd1} K. Glampedakis and N. Andersson, Class. Quant. Grav. {\bf 20}, 3441 (2003).

\bibitem{Det} S. Detweiler, Astrophys. J. {\bf 239}, 292 (1980).

\bibitem{GlaAnd2} K. Glampedakis and N. Andersson, Phys. Rev. D {\bf
64}, 104021 (2001).

\bibitem{Car} V. Cardoso, Phys. Rev. D {\bf 70}, 127502 (2004).

\bibitem{Hod1} S. Hod, Phys. Rev. D {\bf 75}, 064013 (2007).

\bibitem{Hod2} S. Hod, Class. and Quant. Grav. {\bf 24}, 4235 (2007).

\bibitem{Gruz} A. Gruzinov, arXiv:gr-qc/0705.1725.


\bibitem{Teu} S. A. Teukolsky, Phys. Rev. Lett. {\bf 29}, 1114 (1972);
Astrophys. J. {\bf 185}, 635 (1973).

\bibitem{Heun} A. Ronveaux, {\it Heun's differential equations}.
(Oxford University Press, Oxford, UK, 1995).

\bibitem{Flam} C. Flammer, {\it Spheroidal Wave Functions} (Stanford
University Press, Stanford, 1957).

\bibitem{TeuPre} S. A. Teukolsky and W. H. Press, Astrophys. J.
{\bf 193}, 443 (1974).

\bibitem{StaChu} A. A. Starobinsky, Zh. Eksp. Teor. Fiz. {\bf 64},
48 (1973) [Sov. Phys. JETP {\bf 37}, 28 (1973)]; A. A. Starobinsky
and S. M. Churilov, Zh. Eksp. Teor. Fiz. {\bf 65}, 3 (1973) [Sov.
Phys. JETP {\bf 38}, 1 (1973)]

\bibitem{PreTeu} W. H. Press and S. A. Teukolsky, Astrophys. J.
{\bf 185}, 649 (1973).

\bibitem{Notedelta} Had we taken $\Re\delta<0$ or $\Im\delta<0$, we would have found
that $\epsilon\gg 1$. In this case a consistent solution of the
resonance condition, Eq. (\ref{Eq6}), would require
$1/\Gamma(1/2+2i\hat\omega+i\delta-4i\tau/\sigma)=O(\epsilon^{-1})$.
It is straightforward to show that the black-hole resonances would
still be given by the same analytical expressions, see Eqs.
(\ref{Eq11})-(\ref{Eq12}) below.

\bibitem{Abram} M. Abramowitz and I. A. Stegun, {\it Handbook of
Mathematical Functions} (Dover Publications, New York, 1970).

\bibitem{Notevalid} It is worth emphasizing again that this simple relation
is valid in the $\Im\omega\ll\Re\omega$ regime.

\bibitem{Hodb} S. Hod, Phys. Lett. B {\bf 666} 483 (2008).

\bibitem{Notedim} We note that a spherically symmetric Schwarzschild black hole has
only one time/length scale-- its horizon radius, $r_+$ (or
equivalently, its mass $M$). One therefore expects to find $\tau
\sim r_+$ (and $\omega_I \sim r^{-1}_+$) on dimensional grounds. On
the other hand, rotating Kerr black holes have an additional
lengthscale-- the black-hole inverse temperature $T^{-1}_{BH}$. Here
we have established that the relevant relaxation timescale of a
perturbed black hole is determined by its inverse temperature,
$T^{-1}_{BH}$, and not by its horizon radius $r_+$. We emphasize
that $T^{-1}_{BH}$ is much larger than $r_+$ in the extremal limit,
$T_{BH} \to 0$. Thus, our result $\min\{\Im\omega\}=\pi
T_{BH}+O(T_{BH}^2)$ is {\it stronger} than a relation of the form
$\min\{\Im\omega\}\sim r^{-1}_+$, which one could have anticipated
from some naive dimensionality considerations. In particular, the
present analytical results along with numerical computations
\cite{Leaver,Ono,GlaAnd2int,GlaAnd1} imply that extremal Kerr black
holes have {\it infinitely} long relaxation times.

\end{thebibliography}
\end{document}